\documentclass{gpSTS}
\usepackage{setspace}
\title{Autonomous Investigations over WS$_2$ and Au\{111\} with Scanning Probe Microscopy}

\author[1*]{John C. Thomas}
\author[1,2]{Antonio Rossi}
\author[3]{Darian Smalley}
\author[1]{Luca Francaviglia}
\author[4,5]{Zhuohang Yu}
\author[4,5]{Tianyi Zhang}
\author[4,5]{Shalini Kumari}
\author[4,5]{Joshua A. Robinson}
\author[4,5,6]{Mauricio Terrones}
\author[3]{Masahiro Ishigami}
\author[2]{Eli Rotenberg} 
\author[1]{Edward S. Barnard}
\author[1]{Archana Raja}
\author[1]{Ed Wong}
\author[1]{D. Frank Ogletree}
\author[7*]{Marcus M. Noack}
\author[1*]{Alexander Weber-Bargioni}
\affil[1]{Molecular Foundry, Lawrence Berkeley National Laboratory, Berkeley, CA 94720, United States of America}
\affil[2]{Advanced Light Source, Lawrence Berkeley National Laboratory, Berkeley, CA 94720, United States of America}
\affil[3]{Department of Physics and NanoScience Technology Center, University of Central Florida, Orlando, FL 32816, United States of America}
\affil[4]{Department of Materials Science and Engineering, The Pennsylvania State University, University Park, PA 16082 United States of America}
\affil[5]{Center for Two-Dimensional and Layered Materials, The Pennsylvania State University, University Park, PA, 16802 United States of America}
\affil[6]{Department of Physics and Department of Chemistry, The Pennsylvania State University, University Park, PA, 16802 United States of America}
\affil[7]{Center for Advanced Mathematics for Energy Research Applications, Lawrence Berkeley National Laboratory, Berkeley, CA 94720, United States of America}
\affil[*]{jthomas@lbl.gov, marcusnoack@lbl.gov, afweber-bargioni@lbl.gov}

\date{}
\begin{document}
\maketitle
\section{Abstract}
Individual atomic defects in 2D materials impact their macroscopic functionality. Correlating the interplay is challenging, however, intelligent hyperspectral scanning tunneling spectroscopy (STS) mapping provides a feasible solution to this technically difficult and time consuming problem. Here, dense spectroscopic volume is collected autonomously \emph{via} Gaussian process regression, where convolutional neural networks are used in tandem for spectral identification. Acquired data enable defect segmentation, and a workflow is provided for machine-driven decision making during experimentation with capability for user customization. We provide a means towards autonomous experimentation for the benefit of both enhanced reproducibility and user-accessibility. Hyperspectral investigations on W$S_2$ sulfur vacancy sites are explored, which is combined with local density of states confirmation on the Au\{111\} herringbone reconstruction. Chalcogen vacancies, pristine W$S_2$, Au face-centered cubic, and Au hexagonal close packed regions are examined and detected by machine learning methods to demonstrate the potential of artificial intelligence for hyperspectral STS mapping.
\section{Introduction}
Two-dimensional (2D) material systems are one of the most sought after solid-state and thin-film structures due to the enormous phase space of functionality, which is driven by atomic- and nano- scale defects that can be advantageously manipulated for single-photon emission, strain engineering, Moiré physics, stacking, and transport.\cite{Stern2022,10.1063.5.0072091,Zhang2022,Lin_2016,10.1146/annurev-physchem-050317-021353,Shimazaki2020,Wang2020} Defective perturbations can alter the effective local landscape and immediately impact macroscopic functionality. Scanning tunneling microscopy (STM), one branch of scanning probe microscopy (SPM), remains fundamental for characterizing and understanding material and surface properties at distances within the atomic-to-nano range, providing information at scale into, \emph{e.g.}, spin-orbit coupling effects within chalcogen vacancies, electrically-driven photon emission of individual defects, and substitutional fingerprinting by measuring the local density of states (LDOS). \cite{Zandvliet2009-gx,Hamers1989-je,Peng2021-ae,Chen1998-jv} Techniques that provide spectroscopic insight, such as STM, are extremely important in correlating defective states with macroscopic phenomena; hyperspectral data collection in, \emph{e.g.}, tip-enhanced Raman imaging and in optical transmission electron microscopy have become standard and enabled spectroscopic capability with enormous information richness that is both spatially and spectrally resolved.\cite{10.1021.nl104163m,10.1080/05704928.2018.1463235,Bannon2009,ophus_2019} However, while hyperspectral STS imaging would provide critical insight into heterogenous electronic properties at the atomic scale, it is not feasible due to the enormous time required. For instance, a hyperspectral optical map collected at 10 min per point in a 150$\times$150 pixel grid would take well over one month. Here, we present a means of performing spatially-dense, point spectroscopic measurements with an STM in combination with artificially intelligent and machine learning (AI/ML) approaches to provide a faster and more reproducible approach to map and identify spectroscopic signatures of heterogeneous surfaces.

Since the inception of scanning tunneling spectroscopy (STS), which measures current-voltage (I/V) characteristics, the vast majority of experiments are performed using single pixel spectroscopy, where routes for data collection tend to be geometrically positioned along a line or grid during point acquisition. The first harmonic, within an atomically resolved area, can be measured over a defined voltage range, with lock-in techniques, that corresponds to a convolution of the tip and surface LDOS (dI/dV). Assuming the tip remains constant then the data collected are from the sample alone,\cite{Novotny2018-vm} where inelastic contributions ($d^2$I/d$V^2$) can also be inferred from the second harmonic.\cite{Stipe1998-jq} This and other recent spectroscopic capabilities within the STM/STS field have given insight into the electronic structure at the atomic scale relevant for the entire field of surface science, chemical processes, optoelectronic processes, the identification of individual adatoms and molecules on surfaces, local spin-orbit coupling, and electron-phonon coupling, all with sufficient spectroscopic energy resolution to also resolve quantum phase transitions, enable the exploration of next generation color centers, the capability to resolve quantum emitters at scale, or map quantum coherence transport.\cite{Lin_2016,Jelic2017-fr,Garg2020-tp,Kimura2021-jh,Zhang2022} En route towards making STS more widely available, we collect point STS autonomously, \emph{via} Gaussian process regression, and benchmark our method on tungsten disulfide (WS$_2$) and a Au\{111\} surface. The methods shown and autonomous experimentation techniques used can be extended to a variety of available spectroscopic techniques within the SPM field, such as force spectroscopy with non-contact atomic force microscopy, however, we first focus on using the LDOS as a tool to identify surface and defective states.

Transition metal dichalcogenides (TMDs), such as WS$_2$, have gained substantial interest for point-defect control,\cite{Barja2019-iq,Wang2018-ag} serving as host substrates for quantum emitters,\cite{Srivastava2015-ya} spin-valley splitting properties,\cite{Schuler2019-bo,Schaibley2016-tw,Schuler2019-vy} and tunable band gap engineering.\cite{Manzeli2017-nr,Liu2014-ki,Li2018-wt} Sulfur vacancies (V$_S$) can be controllably created to serve as target sites for photo- and spin- active functionalization.\cite{Mitterreiter2020-wt,Mitterreiter2021-ih} SPM can measure the electronic characteristics at the atomic level of induced defects,\cite{Barja2019-iq,Schuler2019-bo,Schuler2019-vy} while also providing a path to excite optical transitions.\cite{,Schuler2020-fh} 2D TMDs provide a wide phase space to non-destructively modify the quantum environment through a wide variety of defects,\cite{Schuler2019-bo} however a means to probe the electronic environment to produce statistically significant and reproducible spectra is required to expedite the understanding of emerging phenomena within the field. Furthermore, coinage metal surfaces, such as Au, have been widely explored for a variety of applications in, \emph{e.g.}, molecular self-assembly,\cite{C2CS35365B,ja00019a011} tip-forming,\cite{Schuler2020-fh} and device applications,\cite{Ellis2004,ja063451d} to name a few, and provide a means for tip state calibration with application towards high throughput STS. Hence, W$S_2$ and Au\{111\} are relevant and representative substrates for both the STM/STS community to employ as model systems and to demonstrate our AI/ML driven approach for hyperspectral STS mapping.

One technical challenge of STM/STS is the difficulty of the technique to acquire reproducible, artifact-free tunneling, especially in heterogeneous samples. Acquisition times are governed by multiple hyperparameters, such as voltage range, step size, and dwell time, where spectral collection that is both highly resolved energetically and spatially is confounded by a variety of factors, which can be very time intensive. In conventional STM/STS, point LDOS exploits the full energetic range of interest with high resolution, and a subsequent dI/dV map can then be collected at a specified energy level for high spatial resolution at the cost of greater experimental broadening. Current imaging tunneling spectroscopy (CITS) consists of scanning the tip in $x_1$- and $x_2$- directions within a predefined grid and collecting a high-resolution spectrum at each pixel and can thus visualize spectroscopic nuances spatially, such as band-bending across defective states, that may otherwise be missed.\cite{Asenjo1992-rw, Cochrane2021} CITS takes advantage of both modalities to create a full spectral and spatial picture of a region of interest, but this modality is complicated by any accompanied thermal drift, piezo hysteresis, grid optimization, and time constraints, which can either introduce artifacts in the spectra or limit experimental acquisition. The need for technical approaches that make hyperspectral STS mapping more accessible and user-friendly can provide essential utility into materials discovery and design.

We propose to make use of Gaussian process (GP) regression for hyperspectral data collection,\cite{Noack2021-jj,Melton2020-xi,Ziatdinov2020-vy,Williams_undated-et,Noack2020-vq} which is a well-known method for function approximation and uncertainty quantification. This method refers to a set of function values, where any finite subset of elements have a joint Gaussian distribution. Given some initial input, a Gaussian prior probability density function is learned and then conditioned on data, providing a posterior mean and variance within the model domain, which can be used to make autonomous decisions about optimal point measurements. This and other learning approaches have shown to be useful for hyperspectral image reconstruction,\cite{Ziatdinov2020-vy,Borodinov2019} autonomous synchrotron experiments,\cite{Melton2020-xi} materials discovery,\cite{Noack2020-vq,Nyshadham2019,annurev-matsci-090319-010954} feature extraction,\cite{Vasudevan2021} and in piezoresponse force microscopy,\cite{acsnano.0c10239,Ziatdinov2020-vy} to name a few applications. The promise of autonomously driven experiments with STM come at the benefit of the human operator and can provide industrial application, where a qualified scientist or engineer can initialize an experiment and allow an AI/ML algorithm to complete the workflow.\cite{Brown2020-kv,Kalinin2021-of,Gordon_2020,sciadv.abb6987,azuri2021role,D1NR01109J,ALLDRITT2022108258} 

Here, we present one technical approach to address this challenge to perform hyperspectral STS mapping at defect sites on two different surfaces and demonstrate a) how to perform measurements with reproducible spectra, and b) create statistically significant electronic characterization of the different intrinsic defects that can be found on samples of interest. While this does not enhance sample throughput directly, it allows for samples to be spectroscopically and automatically interrogated in terms of defect diversity and their electronic fingerprint, such that a non-STM expert would have the ability to produce relevant spectroscopic insight after little training. This is carried out with the combination of two machine learning techniques for autonomous experimentation, where a one-dimensional convolutional neural network (1D-CNN) is used to identify obtained spectra that are collected autonomously by a GP to obtain an accurate CITS representation at a rate that is superior to grid collection. Surface maps are obtained for $V_S$ within WS$_2$ and between known surface reconstructions on a Au\{111\} surface. We further summarize our method in a user-friendly and tailorable software package, gpSTS, for public usage.
\section{Results}
\subsection{Hyperspectral STS Mapping \emph{via} Gaussian Process Regression}
An autonomous hyperspectral STM/STS experiment can be initialized over any substrate that is either conducting or semi-conducting. Spatial parameters and tip offsets in both the $x_1$- and $x_2$- direction are defined by an input image (as defined by point locations $x_1$ and $x_2$ with y signal) that is further used in cross-correlation feature tracking (see Supplementary Notes 1-4). At each point defined by the GP, the bias is ramped over a certain voltage range while the tip is held at constant height. Each spectra can then be identified by a 1D-CNN, where class probabilities are computed, and the sum of dI/dV signal intensity is input into the GP for mean and objective function calculation. As the experiment progresses, a proposed measurement is given to the instrument to collect a point at the uncertainty maximum. A workflow summary of an autonomous experiment is presented in Figure 1, where an atomically sharp tip is directed to the next point for STS point acquisition by exploration that ultimately provides a 3D volume of data defined by both \begin{math}I(x_1,x_2,V)\end{math} and \begin{math}dI(x_1,x_2,dV)\end{math}, where \begin{math}I\end{math} is the current and \begin{math}V\end{math} is the voltage. The true power of this methodology is evident when sufficient orbital information (specific for the V$_S$ deep in-gap state within WS$_2$) and surface structure details (over $Au_{fcc}$ and $Au_{hcp}$) can be obtained in well under 100 collected data points (close to 1\% of the data compared to a CITS grid measurement) that is verified against ground truth data (see Supplementary Figures 1-6) to remove any conceivable bias across experiments, where signal intensity over a given voltage range is monitored.\cite{Schuler2019-vy,Chen1998-jv}    
\begin{figure}[H]
    \centering
    \includegraphics[width = 1 \linewidth]{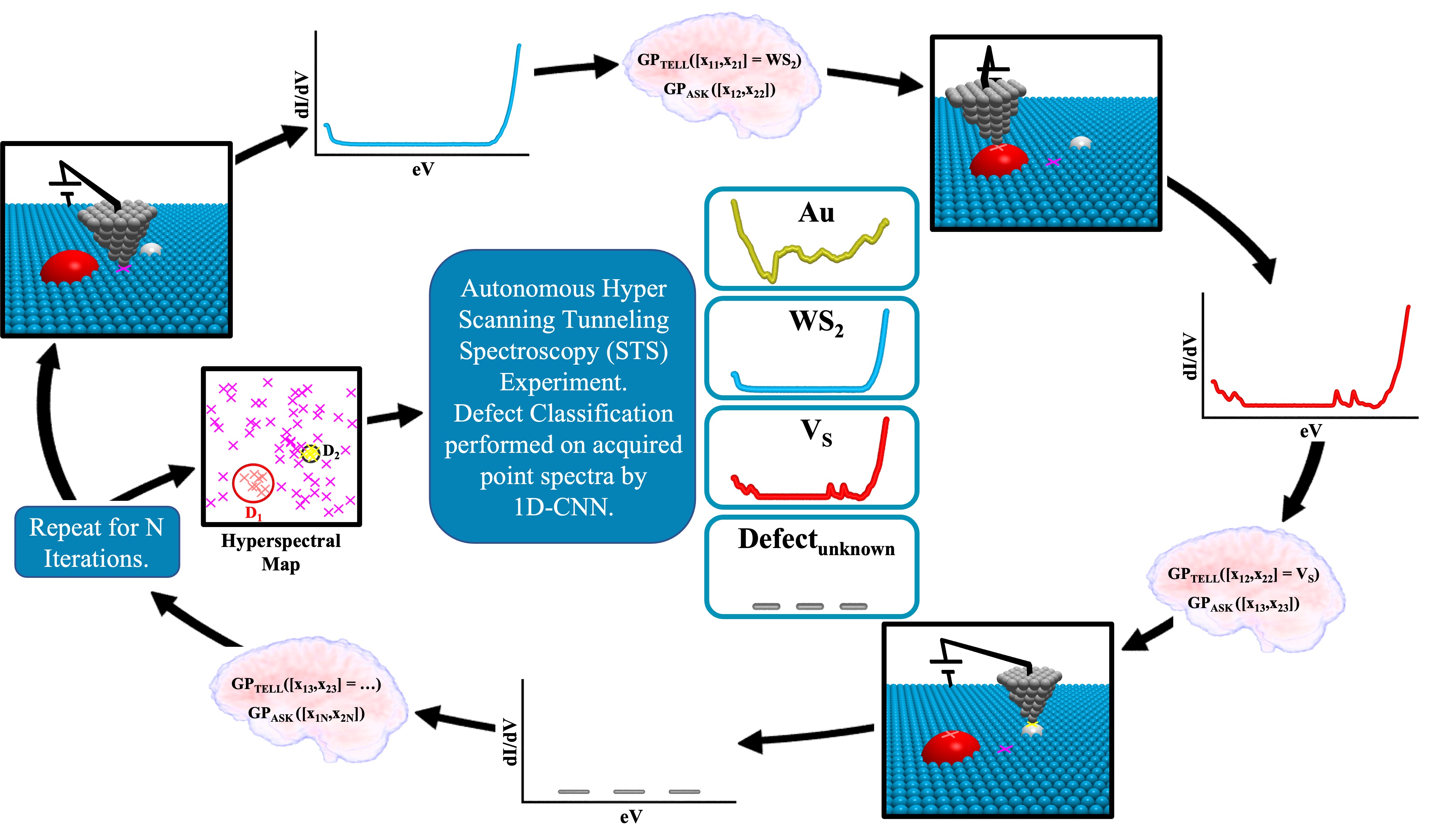}
    \caption{\textbf{Overview of machine-driven hyperspectral STS mapping.} The general workflow of the autonomous experiment performed with a scanning tunneling microscope. The data presented focuses on directing an ultra-sharp metal probe across a given area for point STS acquisition, where both filled and empty states of the sample are examined. After a completed experiment, hyperspectral volume is output from the software and a myriad of substrate and defect classes can be identified, with a trained model, and provided without cognitive bias. The method enables an optimized CITS measurement with predictive capability that was previously inaccessible due to time and tool constraints.}
    \label{fig1:fig_1}
\end{figure}

A GP model can be defined for a given dataset, $D=\left\{\mathbf{x}_i,y_i\right\}$, where the regression model assumes \begin{math}y(x)=f(x)+\epsilon(x)\end{math}, where x are the positions in some input or parameter space, \begin{math}y\end{math} is the associated noisy function evaluation, and \begin{math}\epsilon(x)\end{math} represents the noise term. The variance-covariance matrix \begin{math}\Sigma\end{math} of the prior Gaussian probability distribution is defined via kernel functions \begin{math}k(x_i,x_j;\phi)\end{math}, where \begin{math}\phi\end{math} is a set of hyperparameters that are found by maximizing the marginal log-likelihood of the data (earlier referred to as learning). The Matérn kernel is commonly used to match physical processes and is combined with an anisotropic kernel definition to control the level of differentiability in each direction of the input space.\cite{Williams_undated-et} A predictive mean and variance can then be defined given a Gaussian probability distribution with a set of optimized hyperparameters, which can be further used to find the next optimal point measurements in the GP-driven data acquisition loop (see Supplementary Note 1). GP-driven autonomous experimentation (within the context of Bayesian optimization), where a statistical model of the system is generated based on prior data, uses an acquisition function to suggest the next point of input, which is non-trivial in its design. There are a number of acquisition functions available with different balances of exploration, with the goal to improve the statistical model, and exploitation, with the goal of utilizing the improved statistical model to find the global optimum. 

Here experimental data is collected by exploration, where points are suggested to improve the Gaussian process \emph{via} point selection at uncertainty maxima. The full energy range, which is defined by the accessible voltage range over a given sample that can represent a measurable band gap at both the valence band maximum and conduction band minimum or the range where representative surface states lie (as is the case for $WS_2$ and Au, respectively), can be measured at each point, and indeed we can zoom into any voltage range to visualize orbital information and are able to obtain an optimized hyperspectral map with high resolution both spatially and energetically. In order to evaluate the performance with different user-defined acquisition functions, which either combine posterior mean and variance functions or make use of enabled information-theoretical entities, we perform hyperspectral oversampling on W$S_2$. After an extended and feasibly-obtainable experiment over $\sim$ 30 hours, sufficient data points are collected and we can interpolate over 128$\times$128 pixel grid from acquired data, which is autonomously driven with n = 866 collected data points and shown in Supplementary Figure 1, and compare different acquisition functions using variance, Gaussian process upper confidence bound (GP-UCB), and Shannon’s information gain (SIG) (Supplementary Figures 2-6). Here we use a side-by-side comparison of a GP-driven experiment compared to standard grid methodologies, where the GP point acquisition determined by either variance, GP-UCB, or SIG all out perform comparable grid collections. We additionally show the performance of purely random collected data points (Supplementary Figure 3), where an experiment steered in this fashion shows performance degradation, and consistently and significantly requires more iterations versus a GP-driven CITS measurement (shown over n = 20 experiments). Across acquisition functions, the variance determined from the variance-covariance matrix shows a low of 30 and high of 79 iterations to reach 95\% correlation, the UCB method shows a low of 27 and high of 74 iterations to reach this benchmark, and SIG shows a low of 25 and high of 74 iterations required ($Variance_{iterations}$ = 54.2 ${\pm}$ 14.7, $UCB_{iterations}$ = 53.1 ${\pm}$ 12.4, $SIG_{iterations}$ = 50.0 ${\pm}$ 13.2). Performing a one-way ANOVA analysis, we fail to reject the null hypothesis that the means are equal across acquisition functions, however, we do indeed reject the null hypothesis that means among a random experiment, variance, UCB, and Shannon's information gain are equal ($p_{value}$ << 0.05), which is driven by the elevation and subsequent mismatch of randomly-driven point acquisition.
\subsection{Convolutional Neural Networks for Spectral Identification}
When the sufficient data is collected, we can further successfully identify V$_S$ compared to pristine WS$_2$ and distinguish $Au_{fcc}$ versus $Au_{hcp}$ with a trained 1D-CNN using an 80/20 train/test split ratio on 1482 individually and separately acquired scanning tunneling spectra, consisting of 424 $Au_{fcc}$, 709 $Au_{hcp}$, 158 $V_{S}$, and 191 $WS_2$ spectra (Figure 2). Test data is further split (60/40 ratio) into a validation set used during training to give an estimate of the model's skill and a test set used on unbiased data after training. CNN architectures have shown application towards identifying tip state on H:Si(100),\cite{Rashidi2018-bc} with automated hydrogen lithography,\cite{Rashidi2020-tr,Krull2020-dk} in identifying adatom arrays on a $Co_{3}Sn_{2}S_{2}$ cleaved surface,\cite{Roccapriore2021-qn} and to aid in automating carbon monoxide functionalization with use in noncontact atomic force microscopy.\cite{ALLDRITT2022108258} The CNN architecture chosen uses shared weights to reduce the number of trainable parameters and extract spectral features on the pixel level, which is based on hyperspectral image classification methods used on AVIRIS sensor datasets.\cite{Miclea2020-zh,Ortac2021-ox} 
\begin{figure}[H]
    \centering
    \includegraphics[width = 0.7 \linewidth]{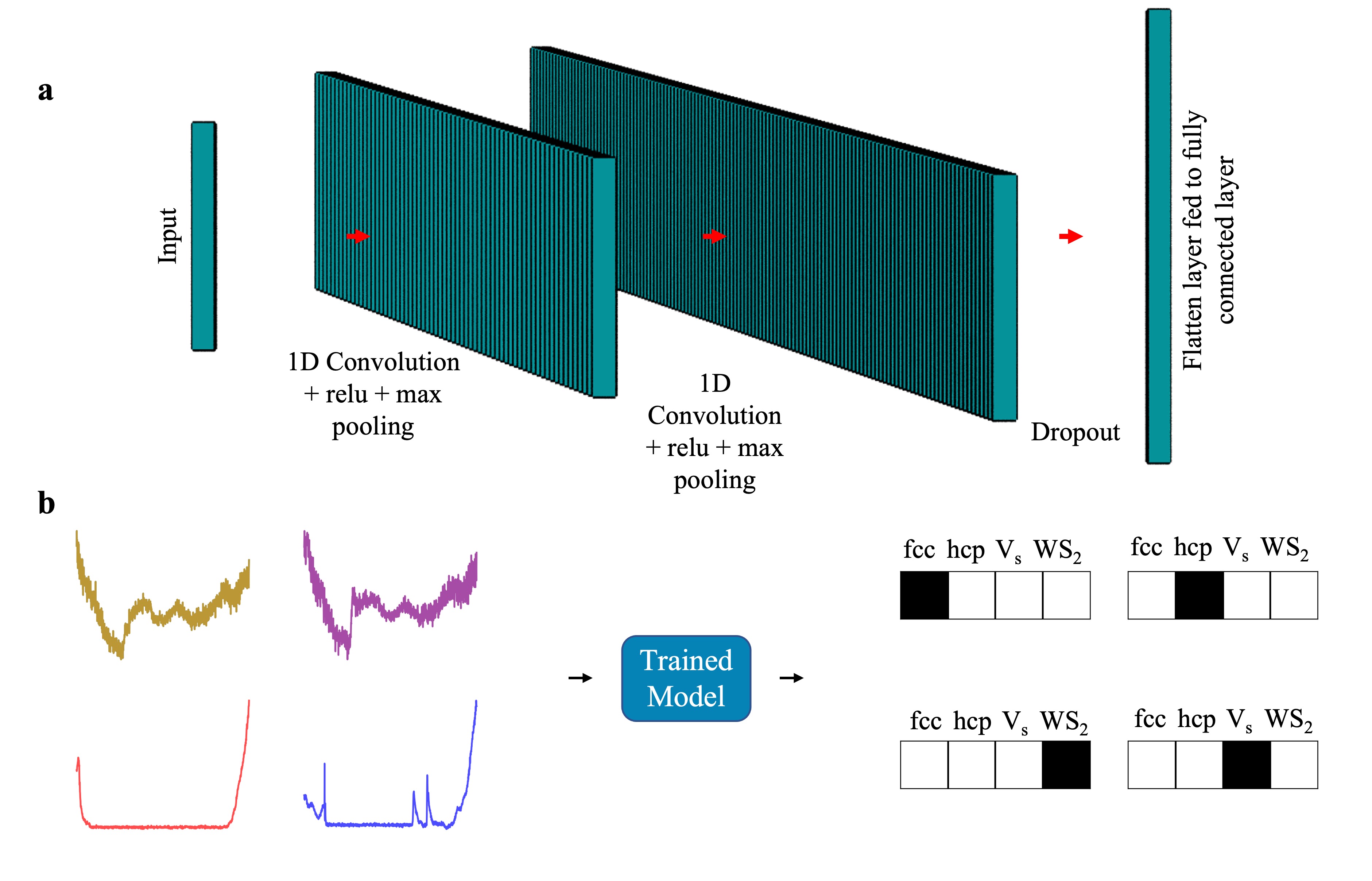}
    \caption{\textbf{1D-Convolutional neural network.} a) Schematic depicting the 1D-CNN model used for training where each layer makes use of a rectified linear unit activation function and max pooling. The first convolutional layer consists of 64 nodes and the second layer makes use of 128 nodes, which is then passed through a dropout layer, flattened, and fed into a fully connected layer. b) Individual spectra can be subsequently passed through a soft-max layer using the trained model to yield class probabilities, where example spectra are shown for $Au_{fcc}$, $Au_{hcp}$, $WS_{2}$, and $V_{S}$. Each spectra depicted exhibits the greatest predictive probability of belonging to the expected class and near zero probabilities for the remaining 3 classes (depicted in gray scale) with the trained model after 6 epochs.}
    \label{fig:fig_2}
\end{figure}

The 1D-CNN contains two convolution layers, one dropout layer to help overcome overfitting, and one fully connected linear layer, where the softmax can then be used after training to obtain point STS class probabilities. Each convolutional layer makes use of a 1x3 kernel to compute the sliding dot product and produce spectral feature maps at each layer (stride 1, padding 1), which is followed by batch normalization, a rectified linear unit (ReLU) activation, and maxpooling layer. The pooling layer down-samples each map while retaining the most important information. ReLU is a nonlinear operation that retains neuron values if it is positive or returns a zero if the input is negative, and is used on both 64 and 128 node layers. During training, the Adam algorithm\cite{Kingma2014-nk} with a learning rate of $10^{-4}$ and computed cross-entropy loss for optimization are used to automatically identify spectral features, where the Adam optimizer minimizes loss. Input and output are shown in Figure 2, where spectra for $Au_{fcc}$, $Au_{hcp}$, $WS_{2}$, and $V_{S}$ that is unseen by the trained model is input and passed through the defined network to produce class identification. Accuracy and loss are further shown through 20 epochs for both training and test data (Supplementary Figure 7), with class accuracy scores presented in Table 1 for the first six epochs. 

\begin{table}[H]
    \centering
    \begin{tabular}{llll}
    \rowcolor[HTML]{C0C0C0} 
    \multicolumn{4}{l}{\cellcolor[HTML]{C0C0C0}Training} \\ \hline
    \rowcolor[HTML]{3166FF} 
    \multicolumn{1}{|l|}{\cellcolor[HTML]{3166FF}$Au_{fcc}$} &
      \multicolumn{1}{l|}{\cellcolor[HTML]{3166FF}$Au_{hcp}$} &
      \multicolumn{1}{l|}{\cellcolor[HTML]{3166FF}$V_S$} &
      \multicolumn{1}{l|}{\cellcolor[HTML]{3166FF}$WS_2$} \\ \hline
    \rowcolor[HTML]{DAE8FC} 
    \multicolumn{1}{|l|}{\cellcolor[HTML]{DAE8FC}63.2} &
      \multicolumn{1}{l|}{\cellcolor[HTML]{DAE8FC}49.5} &
      \multicolumn{1}{l|}{\cellcolor[HTML]{DAE8FC}43.6} &
      \multicolumn{1}{l|}{\cellcolor[HTML]{DAE8FC}64.5} \\ \hline
    \multicolumn{1}{|l|}{98.1} &
      \multicolumn{1}{l|}{98.8} &
      \multicolumn{1}{l|}{93.3} &
      \multicolumn{1}{l|}{97.6} \\ \hline
    \rowcolor[HTML]{DAE8FC} 
    \multicolumn{1}{|l|}{\cellcolor[HTML]{DAE8FC}98.8} &
      \multicolumn{1}{l|}{\cellcolor[HTML]{DAE8FC}99.6} &
      \multicolumn{1}{l|}{\cellcolor[HTML]{DAE8FC}99.4} &
      \multicolumn{1}{l|}{\cellcolor[HTML]{DAE8FC}100.0} \\ \hline
    \multicolumn{1}{|l|}{95.9} &
      \multicolumn{1}{l|}{98.5} &
      \multicolumn{1}{l|}{97.8} &
      \multicolumn{1}{l|}{98.2} \\ \hline
    \rowcolor[HTML]{DAE8FC} 
    \multicolumn{1}{|l|}{\cellcolor[HTML]{DAE8FC}100.0} &
      \multicolumn{1}{l|}{\cellcolor[HTML]{DAE8FC}99.8} &
      \multicolumn{1}{l|}{\cellcolor[HTML]{DAE8FC}100.0} &
      \multicolumn{1}{l|}{\cellcolor[HTML]{DAE8FC}99.5} \\ \hline
    \multicolumn{1}{|l|}{100.0} &
      \multicolumn{1}{l|}{100.0} &
      \multicolumn{1}{l|}{100.0} &
      \multicolumn{1}{l|}{100.0} \\ \hline 
    \rowcolor[HTML]{C0C0C0} 
    \multicolumn{4}{l}{\cellcolor[HTML]{C0C0C0}{\color[HTML]{333333} Validation}} \\
    \rowcolor[HTML]{3166FF} 
    $Au_{fcc}$ &
      $Au_{hcp}$ &
      $V_S$ &
      $WS_2$ \\ \hline
    \rowcolor[HTML]{DAE8FC} 
    \multicolumn{1}{|l|}{\cellcolor[HTML]{DAE8FC}100.0} &
      \multicolumn{1}{l|}{\cellcolor[HTML]{DAE8FC}24.1} &
      \multicolumn{1}{l|}{\cellcolor[HTML]{DAE8FC}0.0} &
      \multicolumn{1}{l|}{\cellcolor[HTML]{DAE8FC}100.0} \\ \hline
    \multicolumn{1}{|l|}{95.6} &
      \multicolumn{1}{l|}{98.9} &
      \multicolumn{1}{l|}{95.7} &
      \multicolumn{1}{l|}{100.0} \\ \hline
    \rowcolor[HTML]{DAE8FC} 
    \multicolumn{1}{|l|}{\cellcolor[HTML]{DAE8FC}100.0} &
      \multicolumn{1}{l|}{\cellcolor[HTML]{DAE8FC}96.6} &
      \multicolumn{1}{l|}{\cellcolor[HTML]{DAE8FC}100.0} &
      \multicolumn{1}{l|}{\cellcolor[HTML]{DAE8FC}100.0} \\ \hline
    \multicolumn{1}{|l|}{88.9} &
      \multicolumn{1}{l|}{100.0} &
      \multicolumn{1}{l|}{100.0} &
      \multicolumn{1}{l|}{100.0} \\ \hline
    \multicolumn{1}{|l|}{\cellcolor[HTML]{DAE8FC}95.6} &
      \multicolumn{1}{l|}{\cellcolor[HTML]{DAE8FC}100.0} &
      \multicolumn{1}{l|}{\cellcolor[HTML]{DAE8FC}100.0} &
      \multicolumn{1}{l|}{\cellcolor[HTML]{DAE8FC}100.0} \\ \hline
    \multicolumn{1}{|l|}{95.6} &
      \multicolumn{1}{l|}{100.0} &
      \multicolumn{1}{l|}{100.0} &
      \multicolumn{1}{l|}{100.0} \\ \hline
    \end{tabular}
    \caption{\label{Table 1}\textbf{Model class performance.} Accuracy scores of the first 6 epochs during training.}
\end{table}

Pristine WS$_2$, $V_{S}$, Au$_{hcp}$, and Au$_{fcc}$ training spectra are all optimized within the first 6 epochs, where the model reaches > 95\% accuracy on Au$_{fcc}$ validation data after 5 epochs and 100\% accuracy for remaining classes. Overall test performance reaches 100\% accuracy after 6 epochs, which is the model chosen for classification (see Supplementary Figures 8-10 for test performance metrics). This paves the way for enabling reproducible STS over surface variations that are distinguishable via bias spectroscopy by providing class probabilities for operators to benchmark against, which can be further expanded to any relevant material. Subsequent identification over herringbone reconstruction is performed (Figure 3), where an impurity is used to track drift during a given cycle, dense hyperspectral data is classified with the 1D-CNN, and image segmentation can be performed with individually classed STS point overlays or pixel-by-pixel using an interpolated form. The peak in dI/dV at -0.48 V shows the tendency for low energy surface-state electrons to localize in Au$_{hcp}$ regions.\cite{Chen1998-jv} A completed experiment over a $V_{S}$ within WS$_2$ is also presented showing defect segmentation using the trained 1D-CNN. As most STM/STS data require high-quality tips and surfaces, the data acquired can be used to verify tip quality on both Au\{111\} and $WS_2$, however this is not explicitly explored within the methods presented.
\begin{figure}[H]
    \centering
    \includegraphics[width = 0.7 \linewidth]{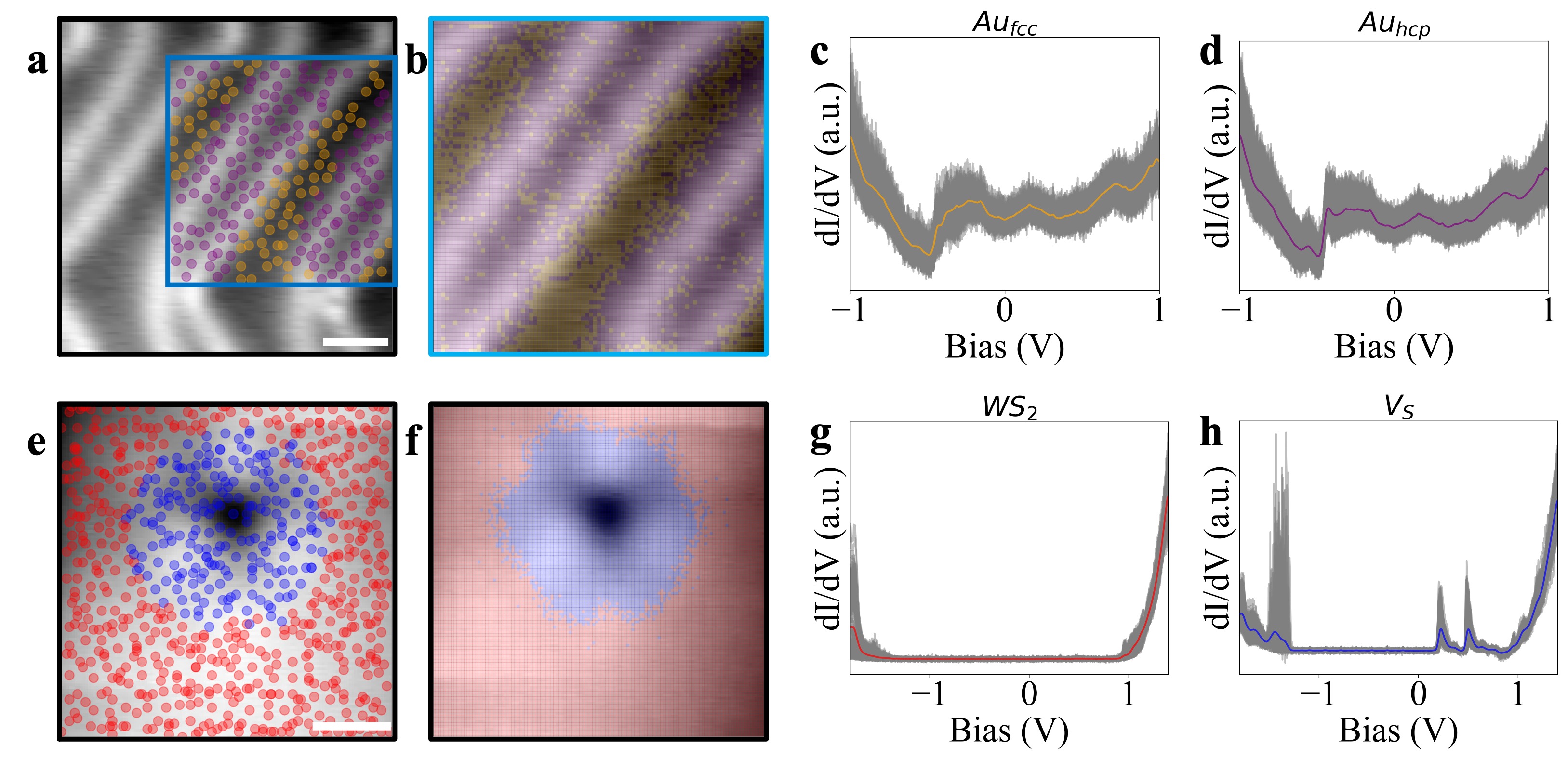}
    \caption{\textbf{Defect identification.} a) Au\{111\} herringbone reconstruction that is identifiable \emph{via} point bias spectroscopy followed by classification using a trained 1D-CNN, where image tracking can be performed on a larger surface region compared to the autonomous STS experiment region. b) Data is further interpolated over a dense grid, classified, and depicted as an image overlay on acquired topography ($I_{tunnel}$ = 30 pA, $V_{sample}$ = 1 V). Scale bar, 2.5 nm. Accumulated spectra over both c) $Au_{fcc}$ and d) $Au_{hcp}$ are shown with the mean spectrum that is colored by classification. e) $V_{S}$ located within \emph{in-situ} annealed $WS_{2}$, where the defect itself is used for drift tracking, with overlaid acquired STS ($I_{tunnel}$ = 30 pA, $V_{sample}$ = 1.2 V) and f) the corresponding linear interpolated form highlighting measured in-gap states. Scale bar, 0.5 nm. Spectra used for training, validation, and test are shown for both g) $WS_{2}$ and h) $V_{S}$, where a total of > 1400 spectra were acquired over multiple experimental runs for both Au and $WS_{2}$.}
    \label{fig:fig_3}
\end{figure}

\subsection{Autonomous Experimentation}

Experiments are performed at liquid helium temperatures and in ultrahigh vacuum to minimize any drift during an experimental run. A number of drift correction techniques have been explored, which take advantage of machine vision techniques, feature tracking, atom-tracking, image pairs, or thermal drift correction methodologies, to list some of the approaches within the literature.\cite{10.1116.1.3360909,OPHUS20161,PhysRevLett.76.459,amantoothpsw,Gaponenko2017} To correct for any residual drift, driven by either thermal fluctuations during piezo motion, sample-to-sample variability, or any tool-to-tool difference, we acquire interval images at a predefined offset window and then compute feature correlation (between spectral acquisition and after n = 10 points) using sliding image patches (Supplementary Figure 11). This block-matching approach is a common technique for image recognition and operates by taking the maximum correlation within a given pixel range.\cite{Thomas2015-ke,Jain1981-dx,Love2006-nz} Any computed offset is registered to the tool by updating the scan window location during the autonomous hyperspectral experiment. Collected images are plane corrected with a line-by-line linear fit to adjust for tilt, since SPM tips are not always perfectly normal to a given sample. Each high resolution spectra is swept from +1.4 V to -1.8 V on WS$_2$ (or +1 to -1 V on Au) and takes 2 minutes for the complete sweep. After two completed autonomous experiments, drift was measured to be on the order of 0.5 ${\pm}$ 0.2 \r{A} on WS$_2$ and 1.1 ${\pm}$ 0.8 \r{A} over Au\{111\} that is shown in Supplementary Figure 12. Hyperparameters can be fine-tuned to best accommodate for any of the defined drift modes during an experiment and for subsequent reproducibility.  
\begin{figure}[H]
    \centering
    \includegraphics[width = 0.7 \linewidth]{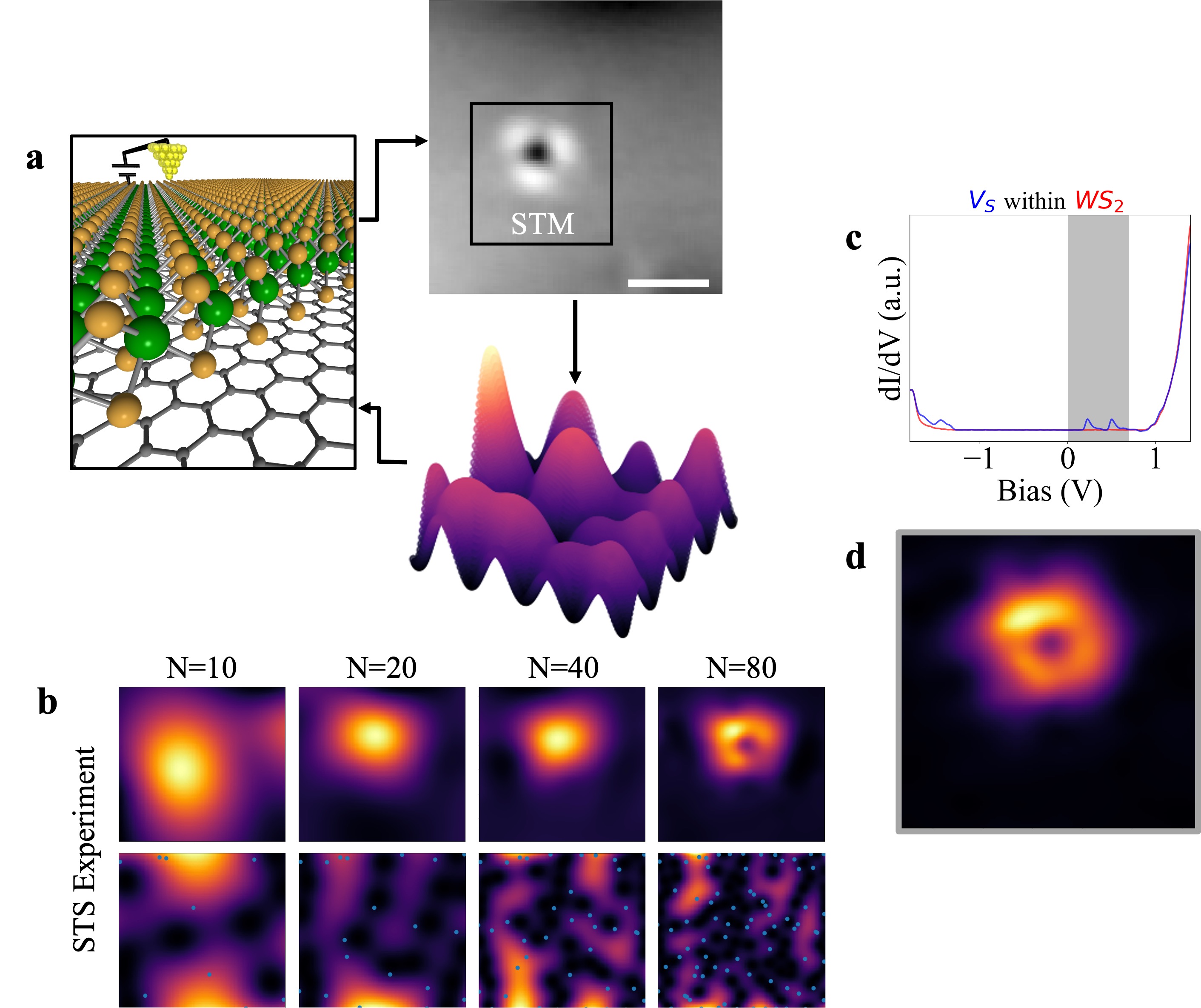}
\caption{\textbf{Gaussian-process-driven experiment.} a) Autonomous scanning tunneling spectroscopy experiment, where an input image ($I_{tunnel}$ = 30 pA, $V_{sample}$ = 1.6 V), showing a $V_S$ within $WS_2$, is used for feature tracking and input into a Gaussian process to determine a corresponding objective and error function to suggest the next point of measurement. Scale bar, 1 nm. b) Evolution of an autonomous experiment showing mean and variance functions at a given interval is depicted, where orbital reconstruction is sufficiently reached with only $\sim$ 1\% of points required compared a 128$\times$128 pixel grid experiment. c) Point defect identification is accomplished at each acquired pixel and the signal summation (0.0 V < $V_{sample}$ < 0.7 V) is used for input. d) Mean model output after N = 160 points, depicting a defect map of measured in-gap W \emph{d} states.}
    \label{fig:fig_1}
\end{figure}

Hyperparameters for both prior-mean model and covariance functions can be optimized after each point acquired during an autonomous experiment. We further show a summary example in Figure 4, depicting the progression of such an experiment in the case of a $V_S$ within WS$_2$, where the user defines the measurement space by providing a topographic image to the software and uncertainty is extracted after every point acquired to determine the next point of acquisition (further detailed in Supplementary Figure 13). The means to direct measurements, optimize a grid, and decrease the time required for sample scrutiny is of key importance within the materials discovery and scanning probe fields, where high-resolution point spectroscopy is on the order of minutes and capturing a dense 128$\times$128 grid with zero drift isn't feasible with available systems (0.14 days with GP compared to 22.76 days for a dense CITS measurement). Enhanced experimental throughput with such a method provides an additional tool for users to collect and identify defects without human bias and/or intervention.

\section{Discussion}
A method for autonomous experimentation is presented that makes use of both GPs and a 1D-CNN for spectral identification, which enables point defect fingerprinting across a wide variety of materials and surfaces. Image segmentation can be subsequently executed after spectral classification. As experiments can be performed without time-intensive input from the operator and at a lower spatial density, high resolution STM/STS can be performed in an autonomous fashion allowing for less redundant information over a given area of interest. Additionally, as neural network algorithms tend to require a large amount of data, the GP can be operated in exploration mode for an increased number of observations to contribute towards statistically significant measurements and ML training on, \emph{e.g.}, an uninvestigated system of interest with spectroscopic signatures not readily available in the literature. The methods presented make use of spectroscopically variant features within a material, where a user can collect data directed by uncertainty quantification or any preferred acquisition function, and use classification to determine tip quality or train on a defined number of classes. We expect that the open source software package can find application across the scanning probe field and greatly increase experimental efficiency, where the library can be easily extended to any system accessible with STM/STS. 

Where previous reports have used either spatial features, decreased pixel density in spectral space, or some combination for segmentation and CITS measurements with an STM, we unambiguously identify defects and surface-state based on high resolution spectral features that leverage the power of Gaussian processes combined with a CNN architecture for prediction. This hyperspectral STS mapping measurement technique that combines CITS with AI/ML enables full characterization of heterogeneous sample surfaces, ensuring that no local spectral features are missed, even by a non-experienced user. 
\section{Methods}
\subsection{Scanning probe microscopy (SPM) measurements}
All measurements were performed with a Createc GmbH scanning probe microscope operating under ultrahigh vacuum (pressure < 2x$10^{-10}$ mbar) at liquid helium temperatures (T < 6 K). Either etched tungsten or platinum iridium tips were used during acquisition. Tip apexes were further shaped by indentations onto a gold substrate. STM images are taken in constant-current mode with a bias applied to the sample. STS measurements were recorded using a lock-in amplifier with a resonance frequency of 683 Hz and a modulation amplitude of 5 mV. 

\subsection{Sample Preparation}
Monolayer islands of WS$_2$ were grown on graphene/SiC substrates with an ambient pressure CVD approach. A graphene/SiC substrate with 10 mg of WO$_3$ powder on top was placed at the center of a quartz tube, and 400 mg of sulfur powder was placed upstream. The furnace was heated to 900 $^{\circ}$C and the sulfur powder was heated to 250 $^{\circ}$C using a heating belt during synthesis. A carrier gas for process throughput was used (Ar gas at 100 sccm) and the growth time was 60 min. The CVD grown $WS_{2}$/MLG/SiC was annealed \emph{in vacuo} at 600 $^{\circ}$C for 30 minutes to induce sulfur vacancies. 

\subsection{Neural Network and Gaussian Process Implementation}
The acquisition software provided leverages the integration of Python and LabVIEW, and makes use of the Nanonis programming interface. The GP was implemented using gpCAM, which is a library for autonomous experimentation by M. M. Noack.\cite{Marcus_gpCAM_2022} The CNN was constructed with Pytorch, which is a deep-learning library available in Python. An Intel Xeon E5-2623 v3 CPU with 8 cores and 64 GB of memory combined with a Tesla K80 with 4992 CUDA cores was used for training.  

\subsection*{Data availability} All data needed to evaluate the conclusions exhibited are present in the paper, supplemental information, and/or available on zenodo (https://doi.org/10.5281/zenodo.5768320). 
\subsection*{Code availability} The full suite of home-built software is available at https://github.com/jthomas03/gpSTS.
\subsection*{Acknowledgments}
This work was supported as part of the Center for Novel Pathways to Quantum Coherence in Materials, an Energy Frontier Research Center funded by the U.S. Department of Energy, Office of Science, Basic Energy Sciences. Work was performed at the Molecular Foundry supported by the Office of Science, Office of Basic Energy Sciences, of the U.S. Department of Energy under contract no. DE-AC02-05CH11231. Work was also funded through the Center for Advanced
Mathematics for Energy Research Applications
(CAMERA), which is jointly funded by the
Advanced Scientific Computing Research (ASCR)
and Basic Energy Sciences (BES) within the
Department of Energy's Office of Science, under Contract No. DE-AC02-05CH11231. S.K and J.A.R. acknowledge support from the National Science Foundation Division of Materials Research (NSF-DMR) under awards 2002651 and 2011839. L.F. acknowledges funding from the Swiss National Science Foundation (SNSF) via Early PostDoc Mobility Grant no. P2ELP2\_184398.
\subsection*{Author Contributions}
J.C.T., A.R., D.S., L.F., M.I., E.R., E.S.B., A.R., M.M.N., and A.W.-B. contributed to data analysis, neural network development, and software implementation. Z.Y., T.Z., S.K., J.A.R., and M.T. synthesized the samples. J.C.T, A.R., E.W., D.F.O., and A.W.-B. helped perform STM measurements and setup experiments. All authors discussed the results and contributed towards the manuscript.  
\subsection*{Competing interests} The authors declare that they have no competing interests. 

\bibliographystyle{ieeetr} 

\end{document}


\maketitle
\section*{Supplementary Notes}
\subsection*{1| Gaussian Process Regression and Acquisition Function}
The GP model can be characterized for a given dataset,
$D=\left\{\mathbf{x}_i,y_i\right\}$, as the probability distribution over functions \begin{math}f(x)\end{math} as
\begin{equation}
p\left(\mathbf{f}\right)=\ -\frac{1}{\sqrt{\left(2\pi\right)^{dim}\left|\mathbf{K}\right|}}exp{\left[-\frac{1}{2}\left(\mathbf{f}-\bm{\mu}\right)^T \mathbf{K}^{-1}\left(\mathbf{f}-\bm{\mu}\right)\right]}
\end{equation}
where, by applying the covariance kernel, \begin{math}K\end{math} is the calculated covariance matrix of \begin{math}D\end{math}, and \bm{$\mu$} is the prior mean vector. The likelihood over \begin{math}\mathbf{y}\end{math} is then
\begin{equation}
p\left(\mathbf{y}\right)=\frac{1}{\sqrt{\left(2\pi\right)^{dim}\left|\mathbf{V}\right|}}\exp{\left[-\frac{1}{{2}}\left(\mathbf{y}-\mathbf{f}\right)^T \mathbf{V}^{-1}\left(\mathbf{y}-\mathbf{f}\right)\right]}
\end{equation}
where \begin{math}\mathbf{V}\end{math} is the matrix of the possibly non-i.i.d. noise (heteroscedastic). From the above equations, the posterior probability distribution for a measurement outcome at \begin{math}\mathbf{x}_0\end{math} can be calculated. The probability distribution mean and variance functions are then defined as
\begin{equation}
m\left(\mathbf{x}_0\right)=\bm{\mu}+\mathbf{k}^T\left(\mathbf{K}+\mathbf{V}\right)^{-1}\left(\mathbf{y}-\bm{\mu}\right)
\end{equation}
\begin{equation}
\sigma^2\left(\mathbf{x}_0\right)=\mathbf{k}\left(\mathbf{x}_0,\mathbf{x}_0\right)-\mathbf{k}^T\left(\mathbf{K}+\mathbf{V}\right)^{-1}\mathbf{k}
\end{equation}
Covariances are computed by the Matérn kernel, which is commonly chosen to match physical processes.\cite{Williams_undated-et} The kernel is defined as
\begin{equation}
k\left(\mathbf{x}_i,\mathbf{x}_j\right)=\frac{2^{1-\nu}}{\Gamma\left(\nu\right)}\left(\frac{\sqrt{2\nu}}{l}r\right)^\nu B_\nu\left(\frac{\sqrt{2\nu}}{l}r\right)
\end{equation}
where \begin{math}r=\parallel{\mathbf{x}_i-\mathbf{x}_j\parallel{}}_{l_2}\end{math}, \begin{math}B_\nu\end{math} is the modified Bessel function, and \begin{math}\nu\end{math} is a parameter defining the differentiability. This is combined with an anisotropic kernel definition to control the level of differentiability in each direction of the input space.

The GP model prior and posterior can be used in the definition of the acquisition function.\cite{Noack2021-jj} The acquisition function can take a variety of forms that include measuring the total uncertainty or favor morphological features. The implemented software can include a user-defined acquisition function that passes through optimization and gives the next point of measurement for data collection. Using the GP defined joint prior given as
\begin{equation}
p\left(\mathbf{f},\mathbf{f}_0\right)=\ -\frac{1}{\sqrt{\left(2\pi\right)^{dim}\left|\Sigma\right|}}\exp{\left[-\frac{1}{2}\left(\begin{matrix}\mathbf{f}-\mu\\\mathbf{f}_0-\mu_0\\\end{matrix}\right)^T\Sigma^{-1}\left(\begin{matrix}\mathbf{f}-\mu\\\mathbf{f}_0-\mu_0\\\end{matrix}\right)\right]}
\end{equation}
where 
\begin{equation}
\Sigma=
\begin{pmatrix}K&\kappa\\\kappa^T&\mathcal{K}\end{pmatrix}
\end{equation}
with \begin{math}\kappa_i=k\left(\phi,x_0,x_i\right),\ \mathcal{K}=\ k\left(\phi,x_0,x_0\right)\end{math},
\begin{math}K_{ij}=\ k\left(\phi,x_i,x_j\right)\end{math}, and \begin{math}x_0\end{math} is the point of interest. The acquisition function can now be defined as 
\begin{equation}
f_{acq}\left(x\right)=f_{acq}\left(m\left(x\right),\sigma^2\left(x\right),\mathcal{K},\Sigma\left(x\right)\right)
\end{equation}
and directly provide where the next measurement point will be taken during an autonomous experiment. The simplest example of the acquisition function is the exploration method given as
\begin{equation}
f_{acq}\left(x\right)=\sigma^2\left(x\right)
\end{equation}
where the posterior variance maximum is the next point of acquisition. The general definition of the acquisition function also allows for Bayesian optimization such as using the upper confidence bound (UCB) method given as 
\begin{equation}
f_{acq}\left(x\right)=m\left(x\right)+c\sqrt{\sigma^2\left(x\right)}
\end{equation}
and information theory methods such as the Shannon-information gain defined by
\begin{equation}
f_{acq}\left(x\right)=Entropy(\mathcal{K})-Entropy(\Sigma\left(x\right))
\end{equation}
which can be further fine-tuned to find regions of interest, e.g. using a reference spectrum or image feature. A cost function can also be included, which defines the cost of a given measurement itself and the cost of moving within a defined parameter space.

Here we compare results from using the diagonal of the variance-covariance matrix, UCB, and using the Shannon-information gain (SIG). Additionally, we also compare results from a dense 128$\times$128 grid, an equally spaced 12$\times$12 grid, and purely randomly derived locations.

\subsection*{2| Drift Tracking}
Drift within scanning probe microscopy techniques can be solved through a variety of methods either during the experiment or data manipulation post acquisition. Piezocreep and thermal drift are the two largest contributors to drift, where piezocreep is a nonlinear effect that can be minimized through correcting for the desired tip position to the actual tip location. However, as the piezos move or scan, some amount of thermal drift is expected. A number of image analysis methodologies that make use of features within a raster-scan image have shown efficacy in tracking drift between multiple images.\cite{amantoothpsw} Atom-tracking, where the STM tip is locked onto an atom or molecule using two-dimensional lateral feedback can maintain spatial coordinates and be used to track drift velocity.\cite{PhysRevLett.76.459} Another method that takes advantage of matching sets of key points across two images enables correction of nonlinearities within piezo positioners.\cite{Gaponenko2017} The concept of using image pairs is used across scanning probe microscopy and is also applicable to scanning transmission electron microscopy, where drift distortion can be corrected line-by-line by rectifying scan line origins to minimize differences between measured line intensity and the image recorded along an orthogonal direction.\cite{OPHUS20161} Here, we apply one methodology towards tracking drift during point scanning tunneling spectroscopy acquisition. 

\subsection*{3| Convolutional Neural Network Performance}
Metrics from our proposed model are shown below, where we look at receiver operator characteristics to determine model performance between positive and negative classes, the confusion matrix to view sensitivity, specificity, the false negative rate, and the false positive rate, and the precision recall curve that depicts the true positive rate and the positive predictive value. In all three metrics, the chosen model performs optimally on the data provided against four classes of spectroscopic data. We compare two models with matching hyperparameters, as detailed in https://github.com/jthomas03/gpSTS, that only differ in the number of epochs that the neural network is trained. At 6 epochs, off-diagonal elements, which represent a type i (false positive) or type ii (false negative) error, within the confusion matrix are minimized.

\subsection*{4| gpSTS Experimental Workflow}
A GP experiment can be started over a given region with some initialization parameters, which is provided by the user in the form of a n$\times$n image containing information for pixel size, experimental boundary conditions, and location. Additional information, such as file locations, experiment name, imaging parameters for drift tracking, voltage range and stepsize, and any high-level neural network hyperparameters are input into the configuration file in gpSTS. An example file is shown below. 

\begin{verbatim}
nanonis_config = {
    "Nanonis_Settings": {
        "File": "gpSTSinit",
        "ExperimentName": "Test Out",
        "Version": "0.0.1",
        "ImageStart": "test_img001.sxm",
        "FolderLocation": "\\gpSTS\\src\\",
        "DataLocation": "\\gpSTS\\src\\data\\",
        "Channel": "Z",
        "ImDirection": "forward",
        "SpectralRange": [-1,1],
        "NumSpectralPoints": 1200,
        "Center_Point": [174,34],
        "Search_Window": 40,
        "Feature_Window": 20,
        "ScanCurrent": 30e-12,
        "SpecCurrent": 200e-12,
        "STSbias": "Bias calc (V)",
        "STSsignal": "Current (A)"
    },
    "Neural_Network": {
        "TrainingPath": "\\gpSTS\\src\\train\\",
        "EpochNumber": 2,
        "ClassNumber": 4,
        "LearningRate": 0.001,
        "BatchSizeTrain": 5,
        "BatchSizeVal": 1,
        "BatchSizeTest": 1
    }
}
\end{verbatim}
Acquisition functions choice can also be modified in the configuration file and customized by the user. For more advanced customization, please refer to the documentation for the library for autonomous that is used in gpSTS, gpCAM.\cite{Noack2021-jj}
\begin{verbatim}
###acquisition functions###
def my_ac_func(x,obj):
  mean = obj.posterior_mean(x)["f(x)"]
  cov  = obj.posterior_covariance(x)["v(x)"]
  sig = obj.shannon_information_gain(x)["sig"]
  ucb = mean + 3.0 * np.sqrt(cov)
  return cov
\end{verbatim}
The experiment is then run via command line interface with the Nanonis controller running with the option presented in the article shown below (using the signal summation in acquired spectra).

\section*{Supplementary Figures}

\renewcommand{\figurename}{Supplementary Figure}
\begin{figure}[H]
    \centering
    \includegraphics[width = 0.5 \linewidth]{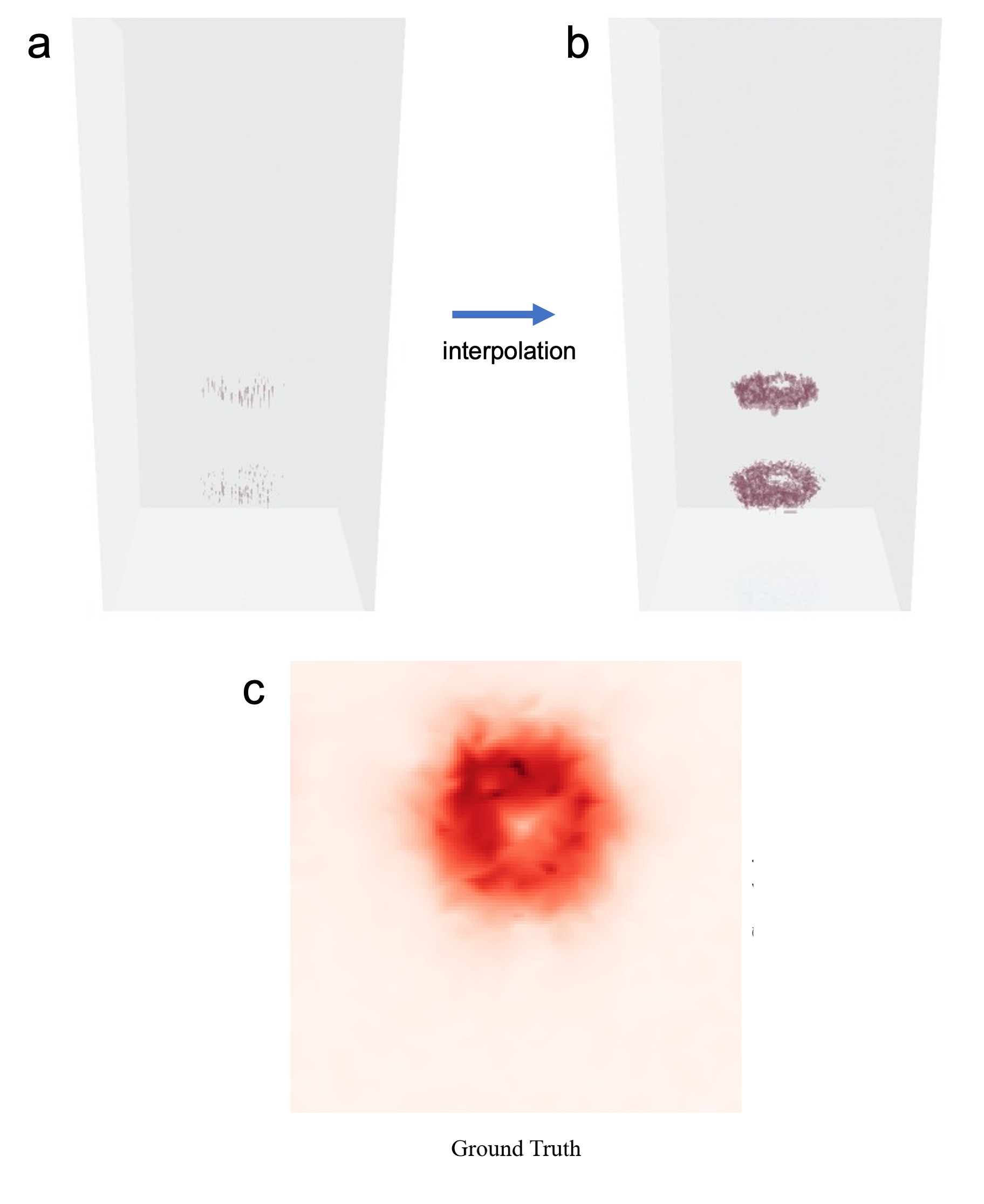}
    \caption{\textbf{Extended Gaussian-process-driven experiment.} a) Ground truth data was formulated from a live experimental run with collected points shown in $x_1$ $\times$ $x_2$ $\times$ $bias_{sample}$ ($V$) space with spectral intensity ($dI/dV$ $a.u.$) depicted in a white to red colorscale, and then b) interpolating the data to a dense 128 x 128 datacube. c) Each point can be summed within a given range (0 to 0.7 V to capture in-gap states), which is shown as an image, and can then be used to compare different acquisition function performance while GP takes point STS shown in b and performs an equivalent summation to reach a predicted mean value function.}
    \label{fig:fig_S1}
\end{figure}

\begin{figure}[H]
    \centering
    \includegraphics[width = 0.7 \linewidth]{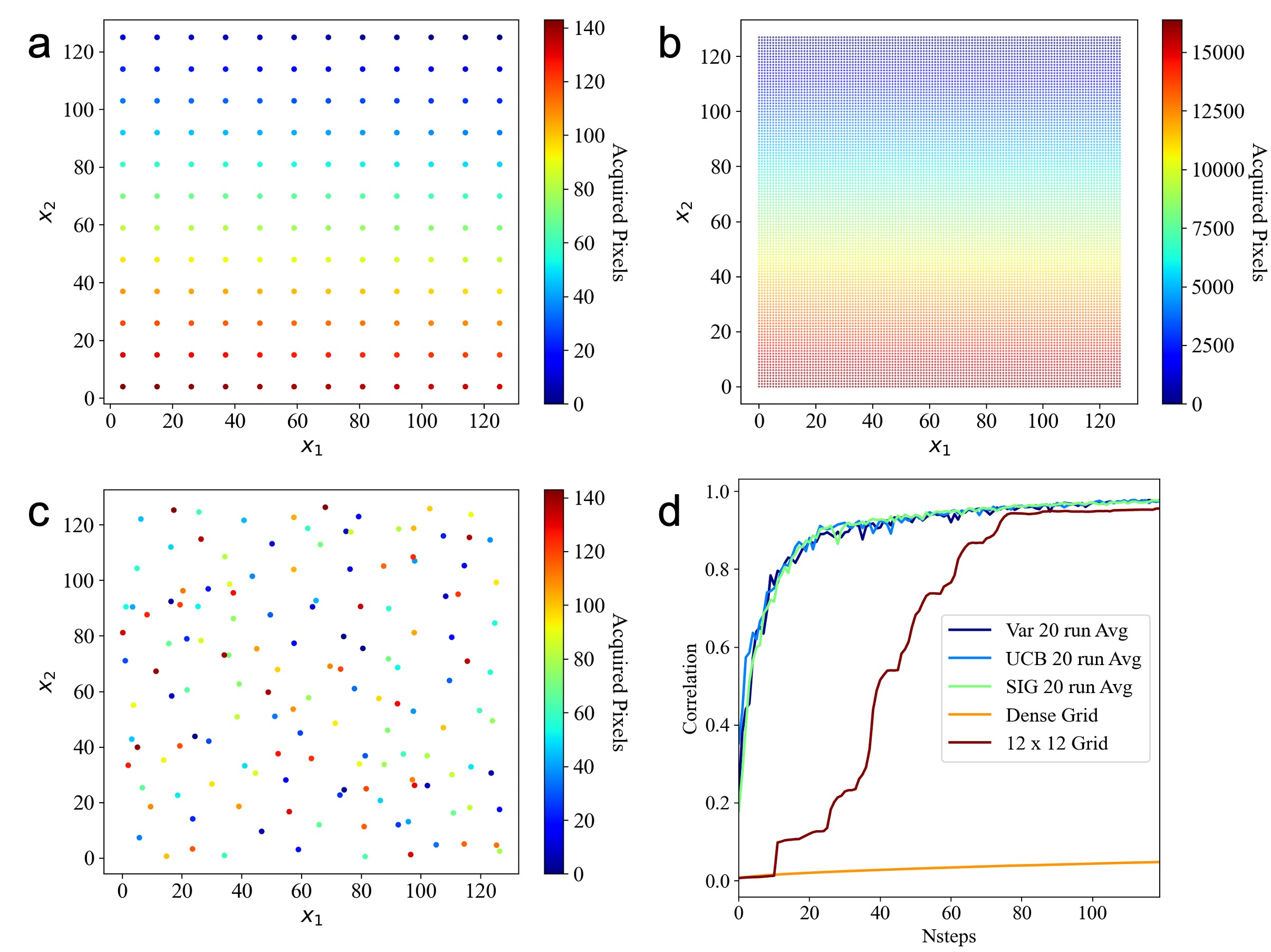}
    \caption{\textbf{Comparison of point collection methods.} In order to test the performance of a GP on low temperature STS data, we input points along two types of grids, compute summation values, interpolate with points acquired, and fill the remaining values with prior STS minima. We show an a) equally spaced 12 x 12 grid, b) a dense 128 x 128 grid, and c) an example GP-driven grid in addition to d) correlations with ground truth data using variance, UCB, Shannon information gain, and each style of grid collection. GP-driven collection shows a measurable boost in correlation and typically reaches greater than 90 percent in under 30 points.}
    \label{fig:fig_S2}
\end{figure}

\begin{figure}[H]
    \centering
    \includegraphics[width = 0.7 \linewidth]{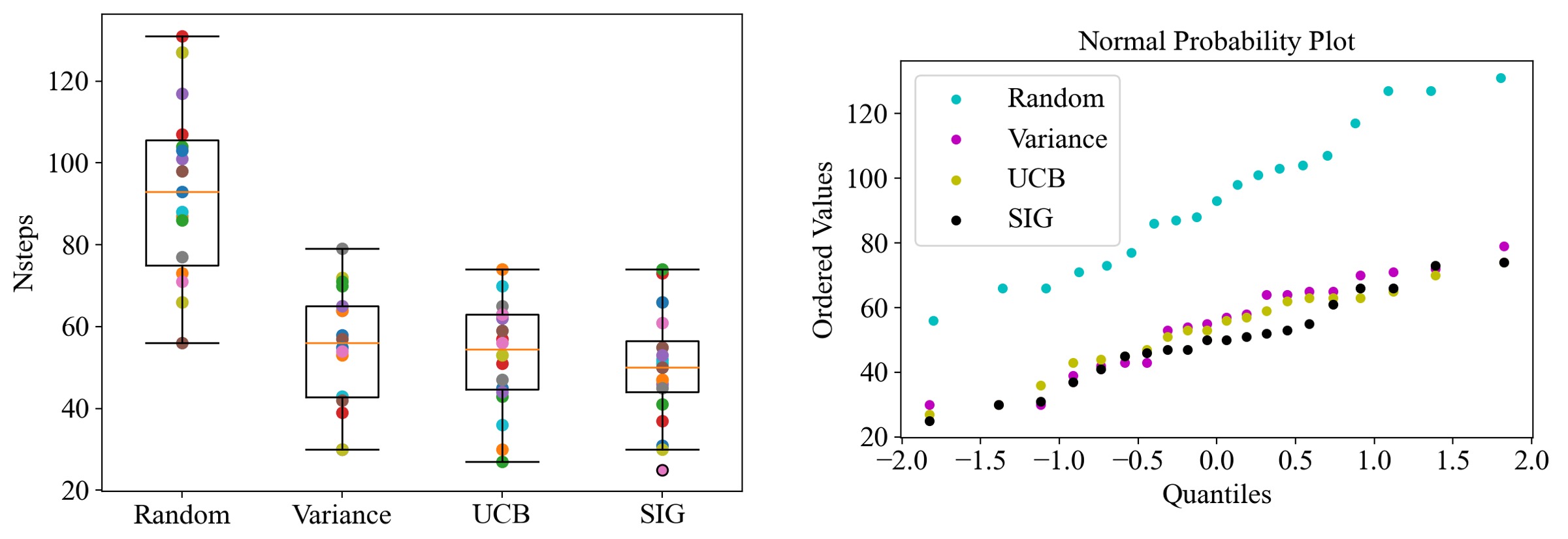}
    \caption{\textbf{Statistical analysis for randomly-driven experimental comparison.} One-way ANOVA depicting number of iterations required to reach 95 percent correlation between a randomized experiment, variance, UCB, and SIG. Here the null hypothesis that mean values are equal is rejected ($p_{value}$ = 3.5 x $10^{-13}$). Comparing across acquisition functions alone, we fail to reject the null hypothesis over 20 runs ($p_{value}$ = 0.61).}
    \label{fig:fig_S3}
\end{figure}

\begin{figure}[H]
    \centering
    \includegraphics[width = 0.7 \linewidth]{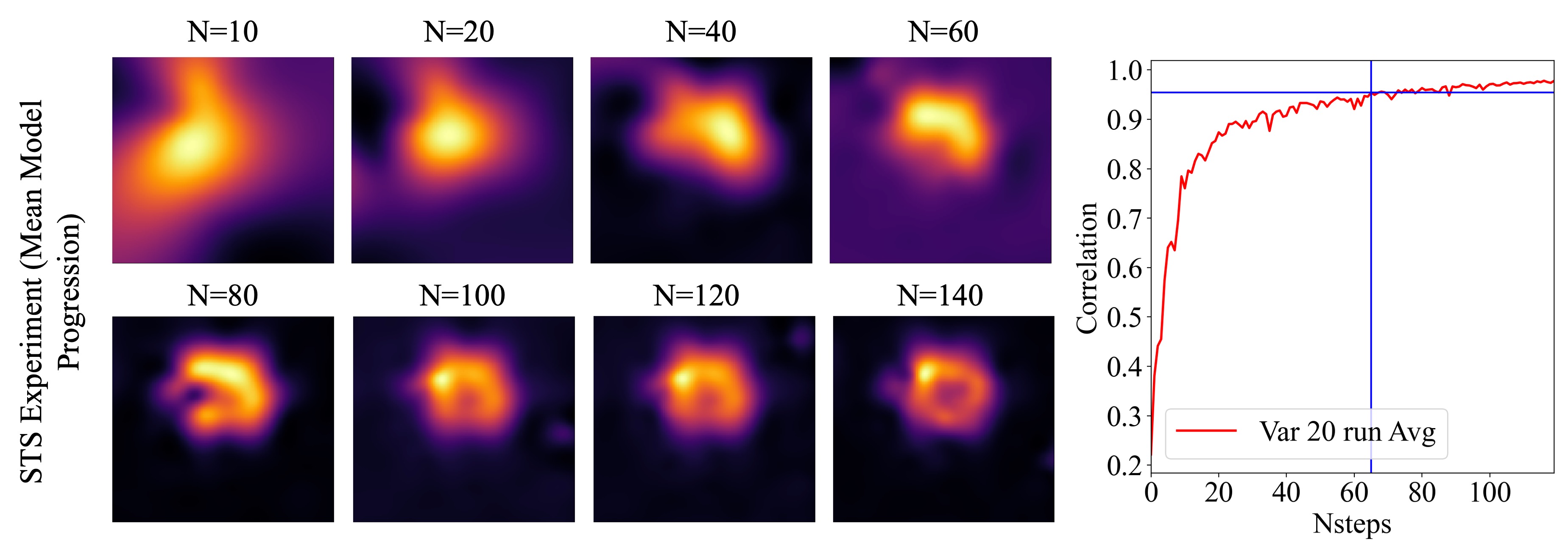}
    \caption{\textbf{Variance.} Correlation reaches 0.954 at 65 intervals with a standard deviation of 0.00987 maintained there after and a maximum correlation of 0.981 reached.}
    \label{fig:fig_S4}
\end{figure}

\begin{figure}[H]
    \centering
    \includegraphics[width = 0.7 \linewidth]{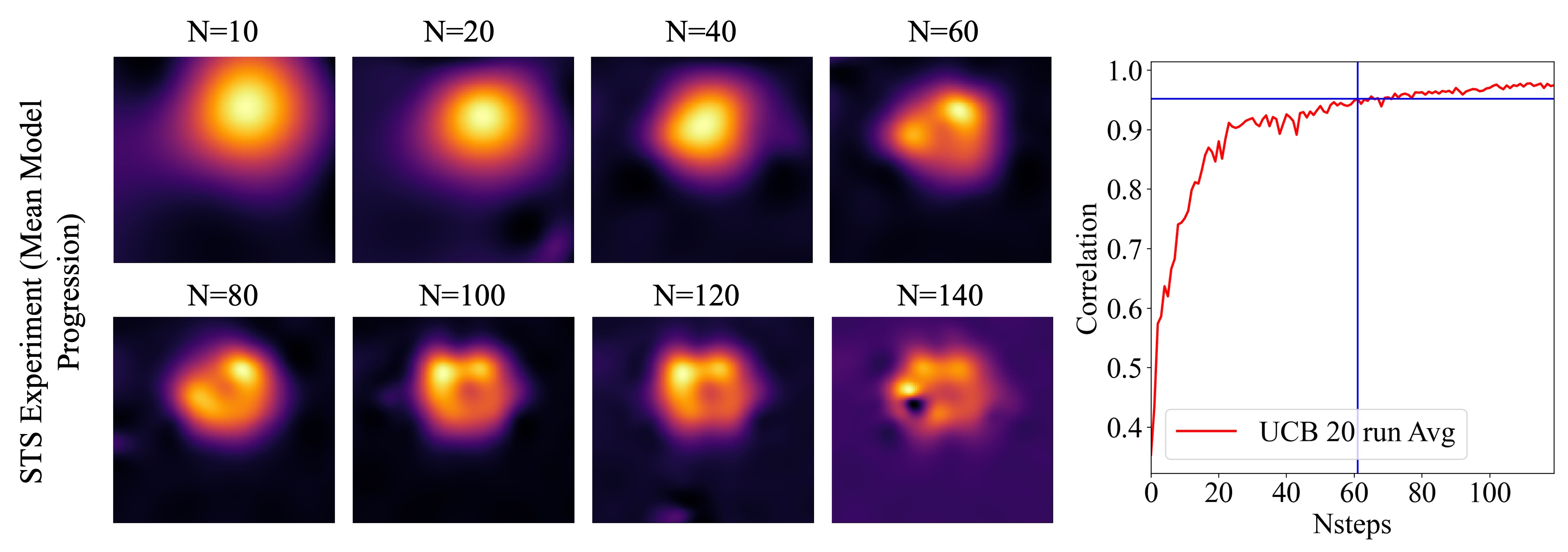}
    \caption{\textbf{UCB.} 0.952 correlation is reached after 61 iterations with a standard deviation of 0.00999 maintained after. A maximum correlation of 0.983 is reached.}
    \label{fig:fig_S5}
\end{figure}

\begin{figure}[H]
    \centering
    \includegraphics[width = 0.7 \linewidth]{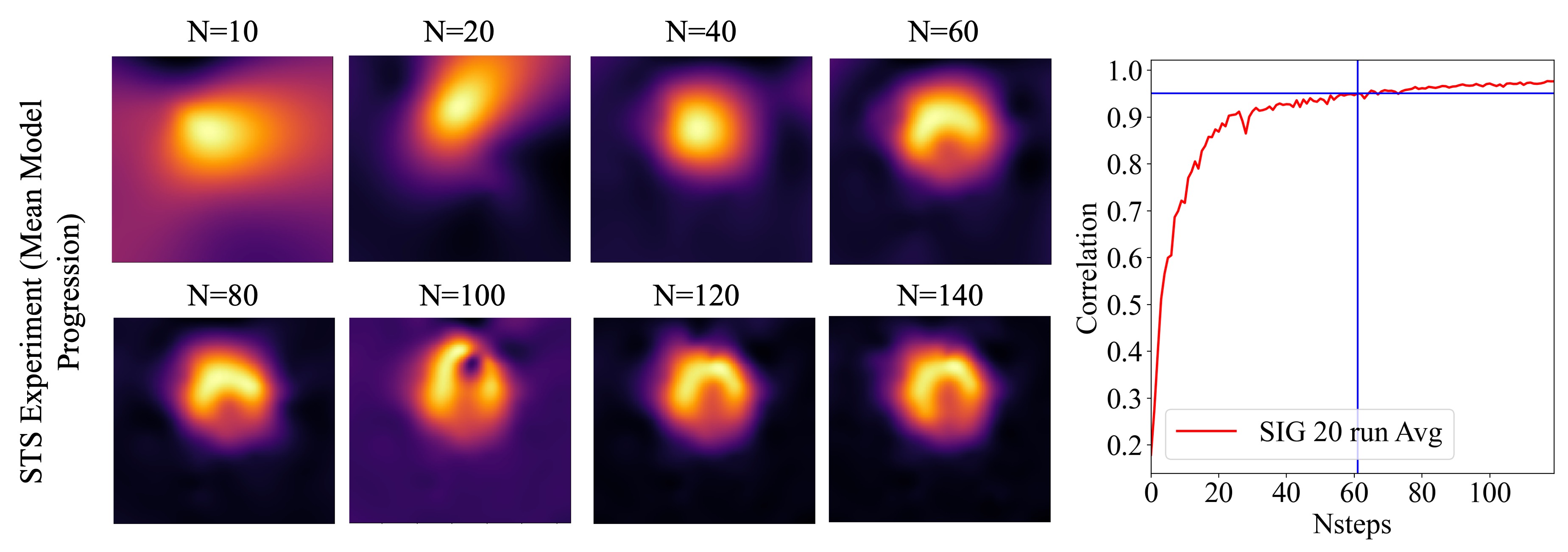}
    \caption{\textbf{Shannon information gain.} 0.951 correlation is obtained at 61 iterations with a standard deviation of 0.00898 maintained after and a maximum correlation of 0.982 reached.}
    \label{fig:fig_S6}
\end{figure}

\begin{figure}[H]
    \centering
    \includegraphics[width = 0.5 \linewidth]{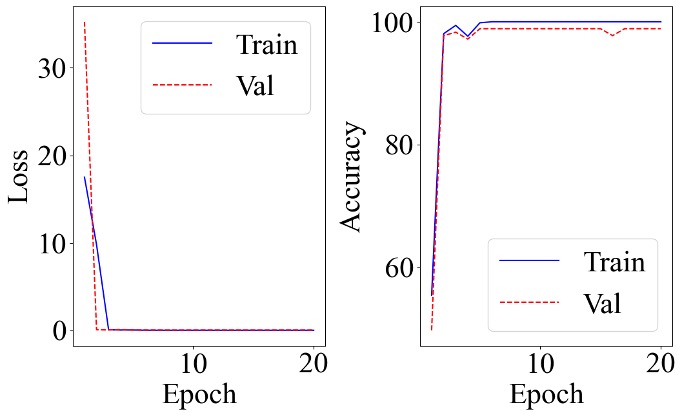}
    \caption{\textbf{1D-CNN model performance.} Accuracy and loss after 20 training epochs is shown for completeness on both training and validation datasets.}
    \label{fig:fig_S9}
\end{figure}
\begin{figure}[H]
    \centering
    \includegraphics[width = 0.7 \linewidth]{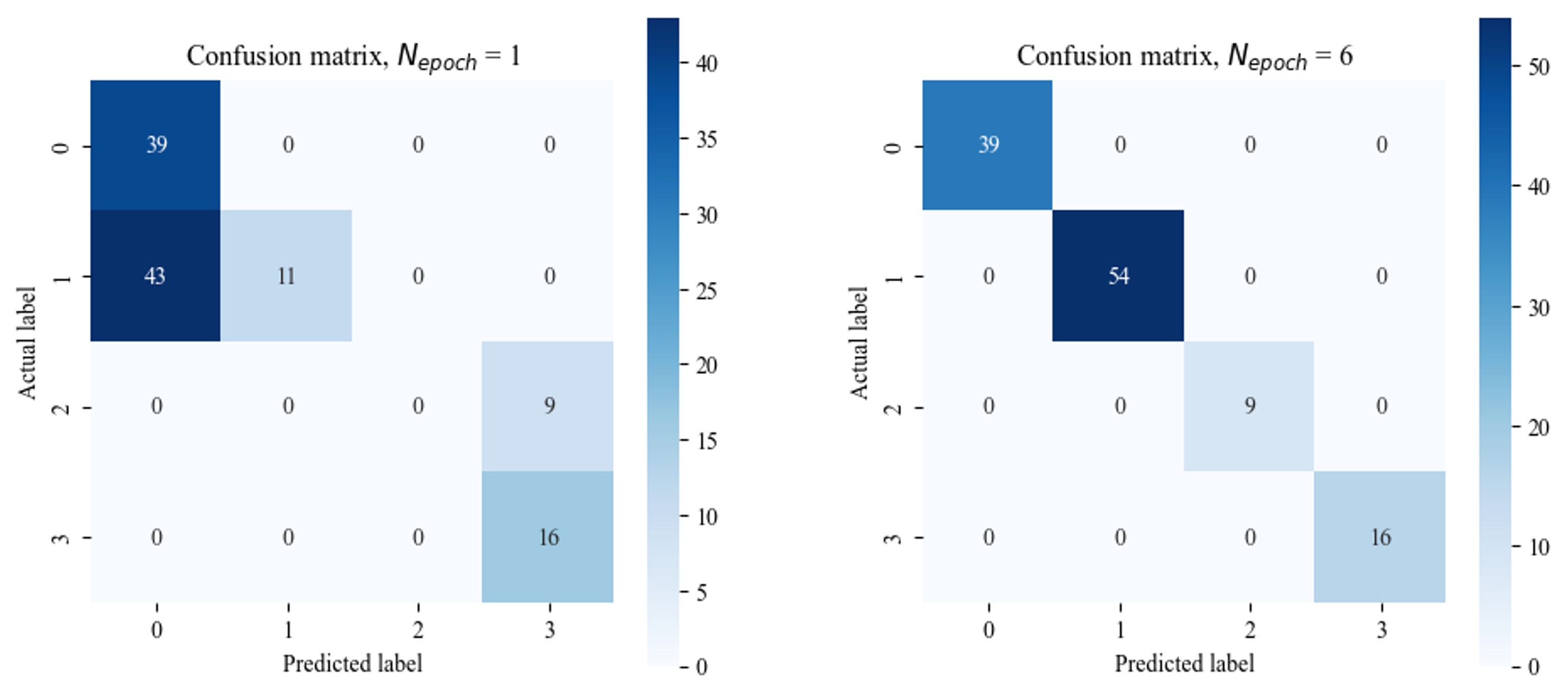}
    \caption{\textbf{Test data classifier error summary.} Confusion matrix taken for classification using the argmax value across probabilities, where error is reduced at 6 epochs.}
    \label{fig:fig_S10}
\end{figure}
\begin{figure}[H]
    \centering
    \includegraphics[width = 0.7 \linewidth]{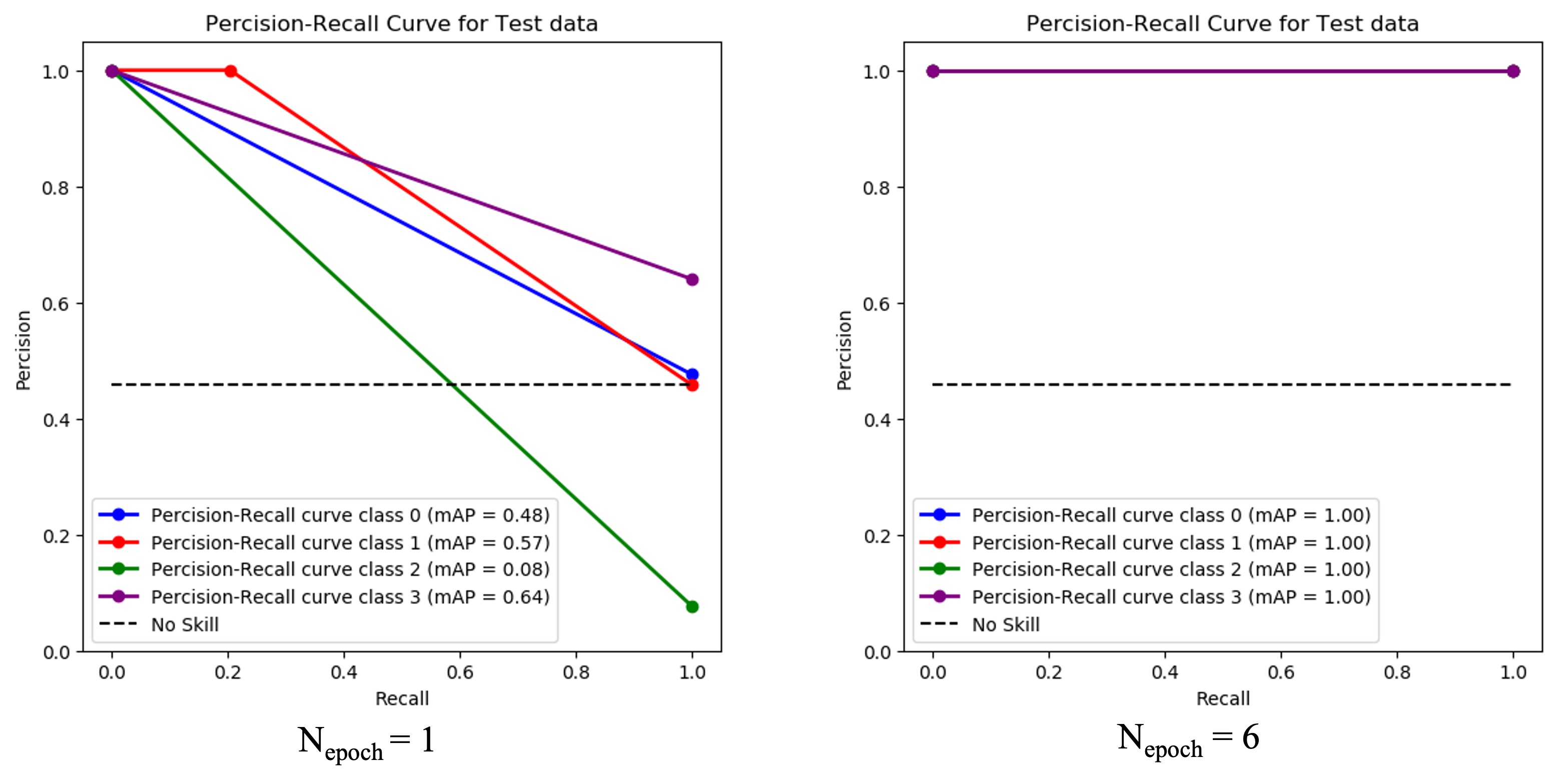}
    \caption{\textbf{Test data precision recall curve.} Shown for test data, where the mean average precision for all classes (class 0 = $Au_{fcc}$, class 1 = $Au_{hcp}$, class 3 = $V_{S}$, class 4 = $WS_{2}$) reaches 1.0 at the end of training.}
    \label{fig:fig_S11}
\end{figure}
\begin{figure}[H]
    \centering
    \includegraphics[width = 0.7 \linewidth]{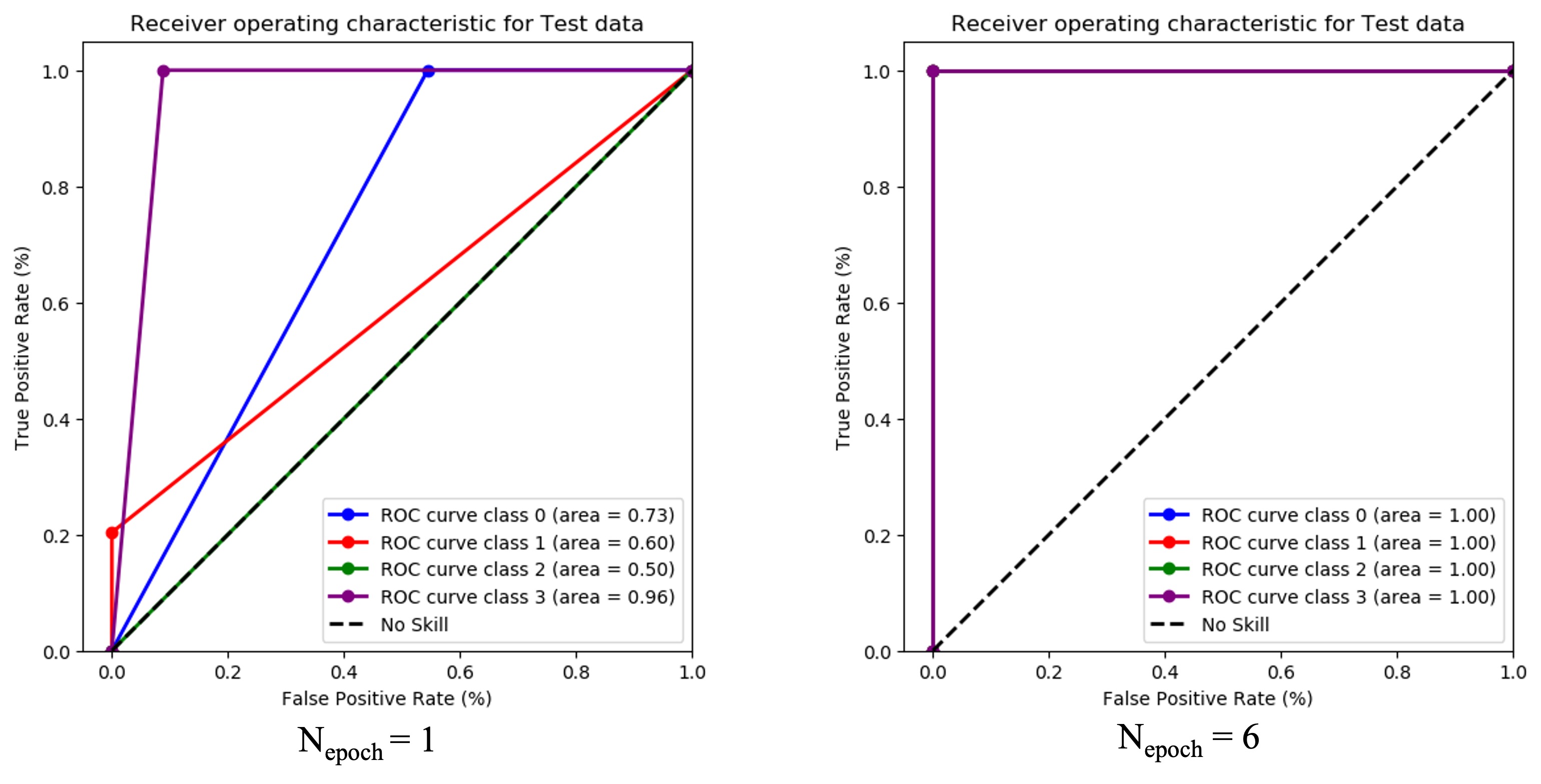}
    \caption{\textbf{Test data ROC curve.} Shown for all classes, where AUC values indicate the model's capability to distinguish between classes after the first and last epoch used in the chosen model.}
    \label{fig:fig_S12}
\end{figure}

\begin{figure}[H]
    \centering
    \includegraphics[width = 0.7 \linewidth]{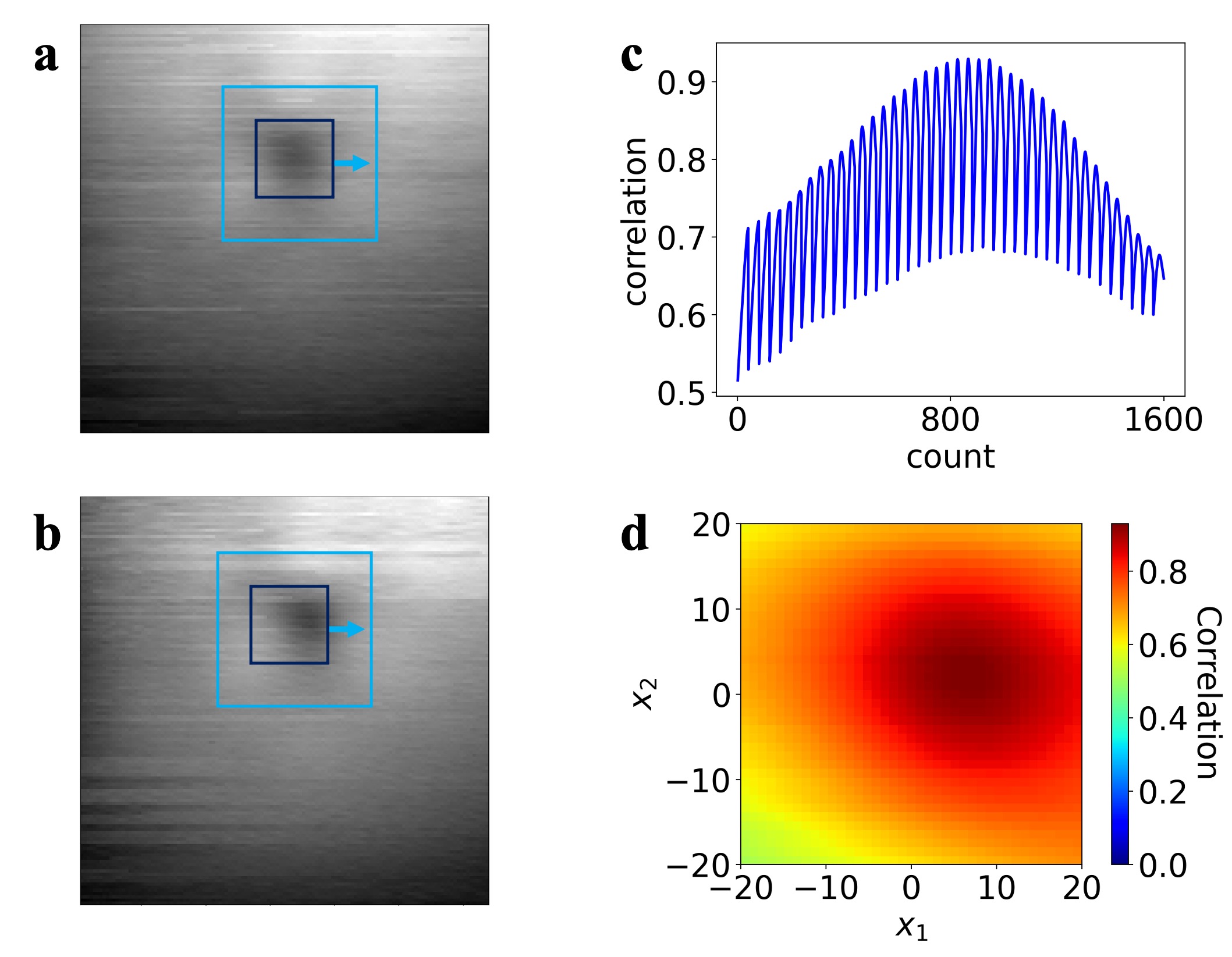}
    \caption{\textbf{Feature correlation.} Cross-correlation over a sequential image taken at a) T = 0 and b) T = 30 min, where correlation is shown in both c) 1D and d) 2D, for clarity. Maximum correlation is obtained at $x_{pixel}$ = 5 and $y_{pixel} = 1$, providing the offset in both the x- and y- directions between collected images.}
    \label{fig:fig_S7}
\end{figure}

\begin{figure}[H]
    \centering
    \includegraphics[width = 0.7 \linewidth]{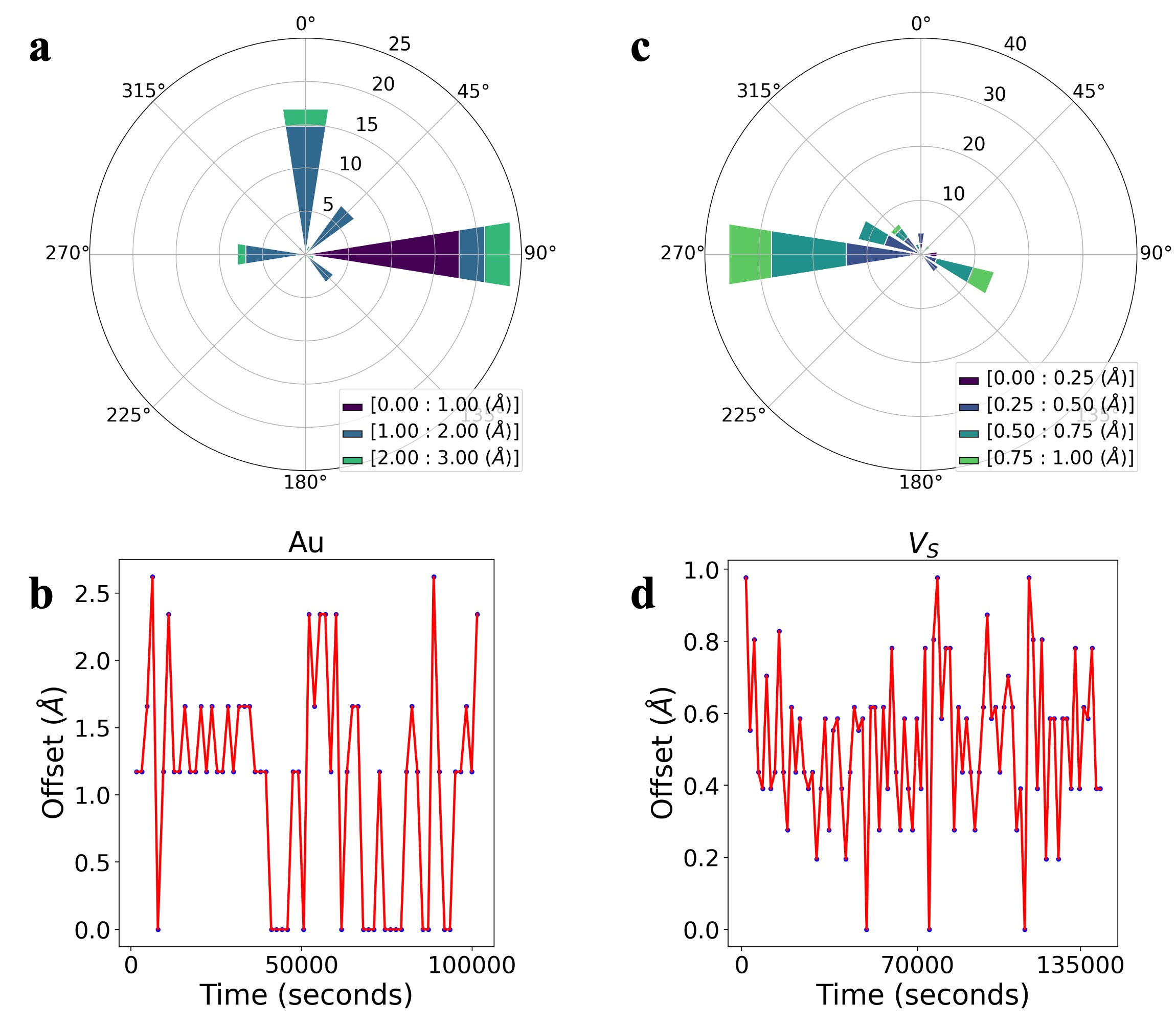}
    \caption{\textbf{Experimental drift.} Computed offsets during a Au\{111\} autonomous experiment is shown a) in two dimensions and b) the magnitude of the 2D vector as a function of time. Both the c) 2D and d) 1D picture are also shown for a $WS_{2}$ experiment. Drift in each example is multi-directional and highlights the need for a means to correct for any offset during measurements, however, parameters can be optimized for any possible tool-tool or run-run delta within the provided software.}
    \label{fig:fig_S8}
\end{figure}

\begin{figure}[H]
    \centering
    \includegraphics[width = 0.7 \linewidth]{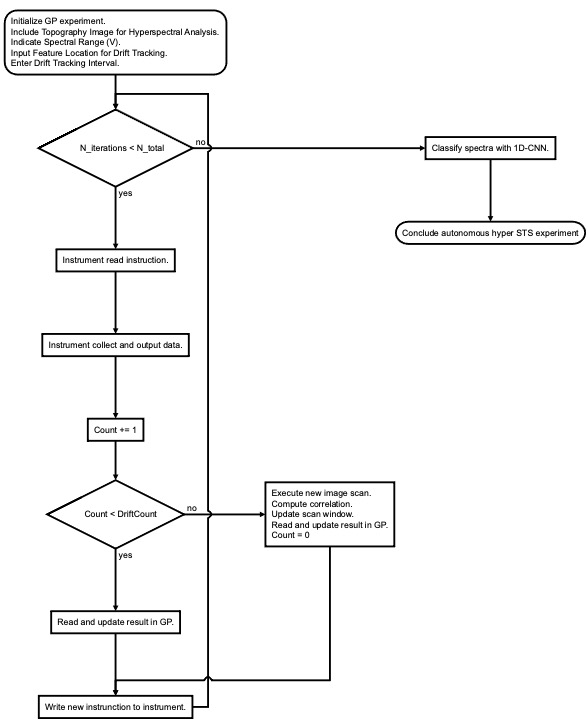}
    \caption{\textbf{Experimental gpSTS flowchart.} Workflow for gpSTS experiment where the signal summation over a given voltage range is used to run the GP, and classification is performed after spectra are acquired. STS are collected in an iterative fashion, and escapes once the number of optimal measurements are obtained.}
    \label{fig:fig_S13}
\end{figure}

\renewcommand\refname{Supplementary References}